\documentstyle[11pt,astron,epsf]{article}

\def\epsout #1 {
  \centering
  \leavevmode\epsffile{./#1}
}

\newcommand{\etal}{{\em et~al.}}

\newcommand{\ionhy}{H{\sc ii}}
\newcommand{\water}{$\mbox{H}_{2}\mbox{O}$}
\newcommand{\methanol}{methanol}

\newcommand{\AIPS}{{\sl AIPS}}

\newcommand{\kms}{$\mbox{km~s}^{-1}$}

\newcommand{\lta}{\raisebox{-0.6ex}{$\,\stackrel
{\raisebox{-.2ex}{$\textstyle <$}}{\sim}\,$}}

\begin{document}
\bibliographystyle{astvar2}

\Large
\begin{center}
{\bf VLBI observations of the 6.7 and 12.2 GHz \methanol\/ masers\\
associated with NGC~6334F} \\[4mm]

\large
S.P.~Ellingsen$^1$, R.P.~Norris$^2$, P.J.~Diamond$^3$, P.M.~McCulloch$^1$,
S.W.~Amy$^2$, A.J.~Beasley$^3$, R.H.~Ferris$^2$, R.G.~Gough$^2$,
E.A.~King$^1$, J.E.J.~Lovell$^1$, J.E.~Reynolds$^2$, A.K.~Tzioumis$^2$
E.R.~Troup$^2$, R.M.~Wark$^2$, M.H.~Wieringa$^2$ \\[2mm]
\end{center}

\normalsize
\noindent
1 Physics Department, University of Tasmania, GPO Box 252C, Hobart 7001, TAS \\
2 Australia Telescope National Facility, PO Box 79, Epping 2121, NSW \\
3 National Radio Astronomy Observatory, PO Box O, Socorro, NM 87801, USA

\section*{Introduction}

The relatively recent discovery of \methanol\/ masers at 6.7 and 12.2 GHz
\cite{Ba1987,Me1991b} has doubled the number of strong maser transitions
found in star formation regions.  The complexity of maser pumping schemes
has meant that the full potential of masers as probes of the physical
conditions has yet to be realised.  Despite this, interferometric observations
of OH and \water\/ masers have revealed much  important information about
the formation of massive stars \cite{Ge1981,Bl1992}.  To date few similar
observations of 6.7 and 12.2 GHz \methanol\/ masers have been published.
However, those which have show that in many cases the 6.7 and 12.2 GHz
\methanol\/ masers have a simple linear or curved spatial morphology
\cite{No1988,No1993}.

Although the general distribution of maser spots can be determined by
observations with connected element interferometers (e.g. ATCA or
VLA), the smaller synthesised beam of VLBI observations means that the
relative positions of maser spots can be measured with much greater
accuracy.  In addition, if there are two or more spots that have very
similar velocities within the synthesised beam of a connected element
interferometer then only one spot is observed, at a position which
depends upon the relative flux density of the individual spots.
However, for VLBI observations the size of the synthesised beam is
much smaller than the typical separation of the maser spots, and so
the position of each of the spots can be determined.

\section*{Observations and data processing} \label{sec:vlbi_obs}

We have used VLBI observations involving five Australian antennas to
image the 6.7 and 12.2~GHz \methanol\/ masers associated with
NGC~6334F.  The properties of each of the antennas is summarized in
Table~\ref{tab:vlbi_array}.  The observations were made on 1992 May 29
and 31, between approximately 7 and 21~h UT on each day.  The data
were recorded using the MK~II VLBI format, with 0.5-MHz bandwidth, and
were correlated using the NRAO MK~II correlator located in Soccoro,
New Mexico.  In spectral-line mode the NRAO MK~II correlator processes
one baseline at a time and produces a 96-channel autocorrelation
spectrum for each antenna and a 192-channel complex cross-correlation
spectrum for each baseline.  The resulting velocity resolution (for
uniform weighting) at 6.7 and 12.2~GHz is 0.28 and 0.15~\kms\/,
respectively, and the velocity range 22.5 and 12.3~\kms\/,
respectively.

After correlation, the data were calibrated and processed using the
\AIPS\/ software package.  The 12.2~GHz observations of NGC~6334F
contained an unblended spectral feature which was strong enough to use
for phase referencing.  For the 6.7~GHz observation of NGC~6334F,
neither phase referencing nor self calibration improved the
signal-to-noise ratio of the images, presumably due to the complexity
of the maser emission.  Each of the spectral channels containing maser
emission was imaged and CLEANed with ``natural VLBI'' weighting.  This
weighting scheme applies natural weighting to the fourth root of the
original weights and was used to reduce the difference in weights
between baselines.

The position and flux density of each maser spot was determined by
fitting a two-dimensional Gaussian to the CLEANed image.  The formal
errors of the Gaussian fit for each of the imaging observations were
0.10--0.40~mas for the 6.7~GHz and 0.10--0.30~mas for the 12.2~GHz
\methanol\/ maser images of NGC~6334F, but these values are probably
underestimated.  Fortunately, we have an independent method of
estimating the accuracy to which the positions of the maser spots can
be determined (see below), and we suggest that the accuracy to which
the relative positions of the masers have been determined is better
than 20\% of the synthesised beam.  For the 12.2~GHz observations
where phase referencing was performed the accuracy is probably
significantly better.

\subsection*{Results and Discussion}

\begin{figure}
  \centering 
  \begin{minipage}[t]{0.80\textwidth}
    \epsfxsize=0.99\textwidth 
    \epsout 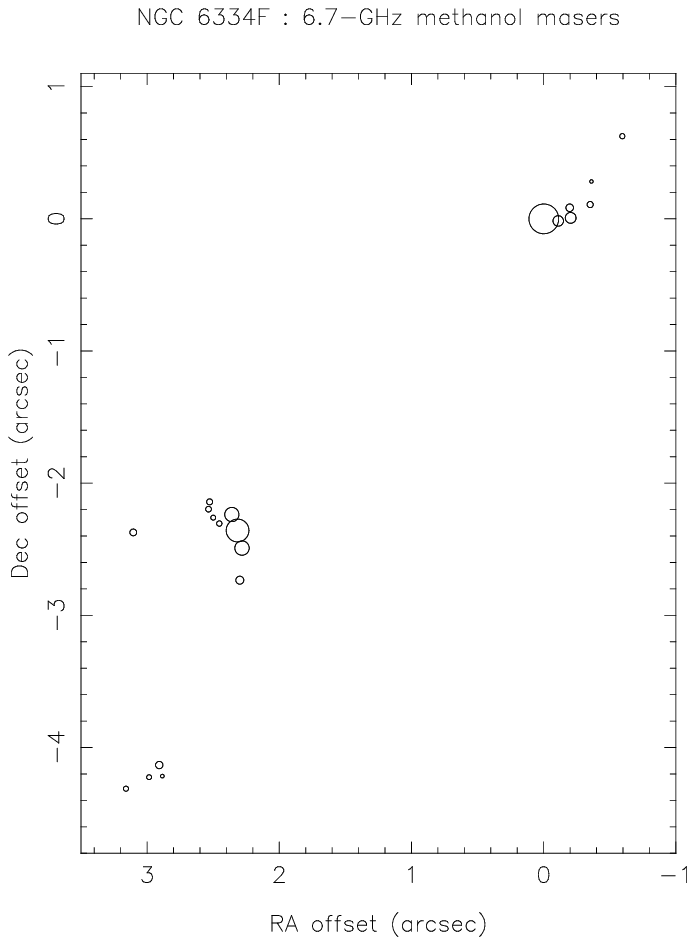 
  \end{minipage}
  \caption[Milli-arcsecond resolution images of the 6.7~GHz
           \methanol\/ masers associated with NGC~6334F]
          {The relative positions of the 6.7~GHz \methanol\/ masers
           associated with NGC~6334F.  The area of the circle marking
           the position of each maser spot is proportional to its flux
           density.  The images were made using a 21.4 {\tt x}
           15.5~mas synthesised beam.}
  \label{fig:ngc6334f_6ghz}
\end{figure}

\begin{figure}
  \centering
  \begin{minipage}[t]{0.80\textwidth}
    \epsfxsize=0.99\textwidth
    \epsout 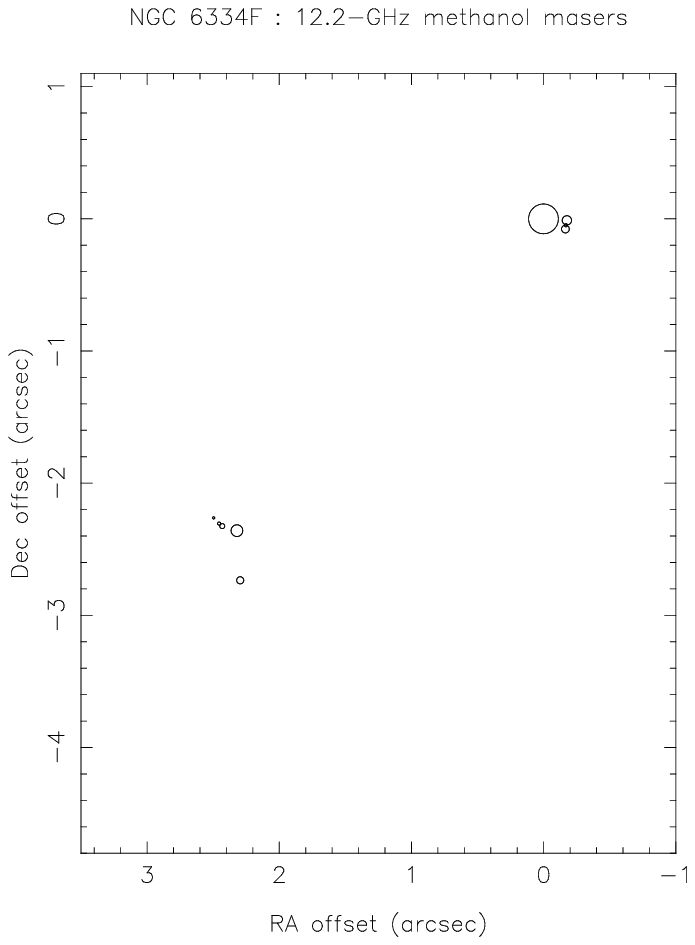
  \end{minipage}
  \caption[Milli-arcsecond resolution images of the 12.2~GHz
           \methanol\/ masers associated with NGC~6334F]
          {The relative positions of the 12.2~GHz \methanol\/ masers 
           associated with NGC~6334F.  The area of the circle marking
           the position of each maser spot is proportional to its
           flux density.  The images were made using a 12.3 {\tt x}
           2.5~mas synthesised beam.}
  \label{fig:ngc6334f_12ghz}
\end{figure}

The compact \ionhy\/ region NGC~6334F is one of the best studied sites
of massive star formation at radio and far-infrared wavelengths
\cite{El1996b,Ga1987,Lo1986b,Mc1979,Ro1982,St1989a}.  It is associated
with some of the strongest known 6.7 and 12.2~GHz \methanol\/ maser
emission.  The 6.7 and 12.2~GHz VLBI images we obtained are shown on
the same scale in Figs~\ref{fig:ngc6334f_6ghz} and
\ref{fig:ngc6334f_12ghz}.  The area of each circle representing a
maser position is proportional to the flux density of that maser spot.
There are three distinct clusters of 6.7~GHz \methanol\/ masers which
we will refer to as NGC~6334F NW (north west), C (central) and S
(south).  At 12.2~GHz only the first two clusters are present and in
each of these there are only approximately half the number of spots
observed at 6.7~GHz.

The 12.2~GHz \methanol\/ masers have previously been imaged by Norris
\etal\/ \cite*{No1988} using the Parkes-Tidbinbilla Interferometer
(PTI).  The 6.7~GHz \methanol\/ masers were also previously imaged by
Norris \etal\/ \cite*{No1993} using the Australia Telescope Compact
Array (ATCA).  The maser distributions we observe agree well with
those of Norris \etal\/ \cite*{No1988,No1993}, with the exception of
the two 6.7~GHz spots that they label C \& E in the central cluster,
which we did not detect.  Either these features are variable and no
longer detectable or, more likely, they result from a blending of
several masers in the larger synthesised beam of the ATCA.  For the
less complicated 12.2~GHz observations the average difference between
my VLBI positions and the PTI positions of Norris \etal\/ is 3~mas,
with the largest offset being only 5~mas.  These two sets of
observations are separated by four years and one month and this
suggests that any proper motion is at a rate of $\lta 0.75$ mas
yr$^{-1}$.  Assuming that the distance to NGC~6334F is 1.7~kpc this
corresponds to a velocity tangential to the line of sight of $\lta$
60~\kms\/.  If the masers are seen edge on in circumstellar discs then the
expected velocity tangential to the line of sight is less than
1~\kms\/, so that this observation does not provide a useful
constraint on models of class~II \methanol\/ masers.  In total, Norris
\etal\/ detected twelve 6.7~GHz and five 12.2~GHz \methanol\/ maser spots,
while our VLBI observations detected twenty 6.7~GHz and nine 12.2~GHz
\methanol\/ maser spots.

The spectral morphology of 6.7- and 12.2~GHz \methanol\/ masers is
often observed to be quite similar \cite{Me1991b,Ca1995a}.  If many of
the individual maser spots at 6.7 and 12.2~GHz have the same
line-of-sight velocities, then it seems logical to assume that they
arise from the same general area of the star formation region.
Observations by Menten \etal\/ \cite*{Me1992} and Norris \etal\/
\cite*{No1993} found the positions of some 6.7- and 12.2~GHz
\methanol\/ masers to be coincident to within the positional errors of
their observations.  As has been noted by Menten \etal\/, this
provides a stringent constraint on any pumping mechanism, as any
scheme which produces 6.7~GHz \methanol\/ masers must also be able to
produce 12.2~GHz \methanol\/ masers and {\em vice versa}.

\begin{figure}
  \centering
  \begin{minipage}[t]{0.80\textwidth}
    \epsfxsize=0.99\textwidth
    \epsout 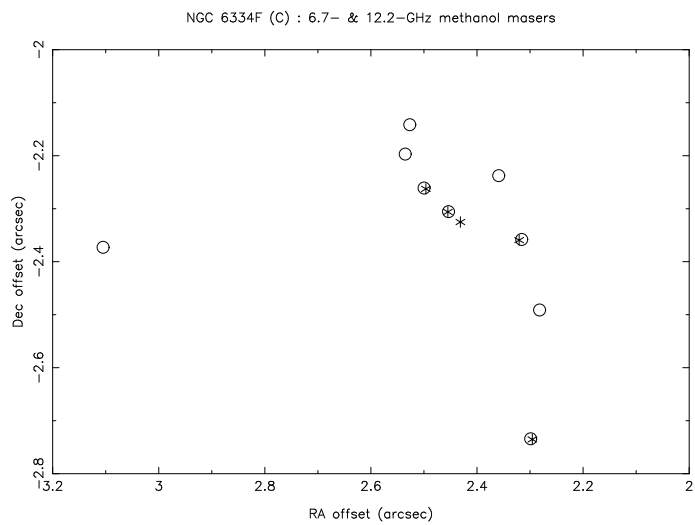
  \end{minipage}
  \caption[6.7- and 12.2~GHz \methanol\/ masers associated with NGC~6334F (C)]
          {The relative positions of the 6.7~GHz \methanol\/ masers
           associated with NGC~6334F (C) are marked by open circles
           and the 12.2~GHz \methanol\/ masers are marked by stars.}
  \label{fig:compare}
\end{figure}

Figs~\ref{fig:ngc6334f_6ghz} and \ref{fig:ngc6334f_12ghz} show the
6.7 and 12.2~GHz \methanol\/ maser emission associated with the
\ionhy\/ region NGC~6334F.  A detailed examination shows that five of the
nine 12.2~GHz \methanol\/ maser spots are coincident (to within
$\approx$ 4~mas) with a 6.7~GHz \methanol\/ maser spot which has the
same velocity (see Table~\ref{tab:comp}).  Figure~\ref{fig:compare}
shows four of the five maser spots which are coincident for the two
transitions.  The rms difference between the positions of the five
masers at 6.7 and 12.2~GHz is 3.7~mas, which is comparable to the
relative positional accuracy.  Thus it appears that the emission from
the two transitions is truly coincident.  This is supported by the
observations of Menten \etal\/ \cite*{Me1992} which had greater
resolution and found positional coincidence to within 1--2~mas.
Further supporting evidence is that the rms difference between the
6.7 and 12.2~GHz positions we observed is dominated by the R.A.
offset (3.6~mas as opposed to 0.7~mas for the Decl.
offset) and the beam shape of these observations was significantly
elongated in R.A.

If we assume that the five maser spots are coincident, then the
observed difference in the positions at the two frequencies will
depend upon the accuracy to which the relative positions of the maser
spots have been estimated by the processing technique outlined above.
As the 6.7~GHz observation did not use phase referencing and had a
larger synthesised beam, it would be expected to have greater errors
in the relative positions of the masers than the 12.2~GHz observation.
If it is assumed that the difference in the positions of the
coincident spots results purely from the errors in relative positions of
the masers at 6.7~GHz then observed rms difference between the 6.7 and
12.2~GHz spots implies errors of 17\% and 5\% of the synthesised beam
for the R.A. and Decl. components respectively.

\section*{Conclusions}

We have made milliarcsecond resolution images of the 6.7- and 12.2~GHz
\methanol\/ maser emission associated with the well-known star
formation region NGC~6334F.  The images agree well with previous lower
resolution observations, but detect approximately double the number of
spots seen in the earlier work.  Comparison of the relative positions
of the 6.7 and 12.2~GHz maser spots shows that five of them are
coincident to within the positional accuracy of these observations
($\approx$ 4~mas).  Menten \etal\/ \cite*{Me1992} observed similar
positional coincidence for W3(OH) and in each case the flux density of
the 6.7~GHz maser spot was greater than that of the 12.2~GHz
\methanol\/ maser spot.  However, for NGC~6334F several of the
coincident maser spots have a larger flux density at 12.2~GHz than at
6.7~GHz.  We also detected several 12.2~GHz \methanol\/ maser spots
with no coincident 6.7~GHz emission.  This implies that, although the
6.7~GHz \methanol\/ masers usually have a greater flux density than
their 12.2~GHz counterparts, regions within the gas cloud exist
where the conditions are more favourable for 12.2 than for 6.7~GHz
\methanol\/ maser emission.  Further milliarcsecond resolution
observations of both 6.7 and 12.2~GHz \methanol\/ masers are required
to determine the distribution of 6.7:12.2~GHz \methanol\/ flux density
ratios.  If the conditions which give rise to the various observed
ratios can be determined, then VLBI images of class II \methanol\/
masers will allow the physical conditions of the star formation region
to be probed with unprecedented resolution.

\newpage

\begin{table}
  \caption[Characteristics of the participating antennas]
          {The diameter, polarization and system equivalent flux density
           characteristics at 6.7 and 12.2~GHz for the 6 antennas
           which participated in the VLBI observations.  RCP = Right
           circular polarization; SEFD = System equivalent flux density}
  \begin{tabular}{lrlrlr} \hline \hline
  {\bf Antenna} & {\bf Diameter} & \multicolumn{2}{c}{{\bf 6.7~GHz}}           &
    \multicolumn{2}{c}{{\bf 12.2~GHz}}          \\
                &                & {\bf Polarization} & {\bf SEFD} &
    {\bf Polarization} & {\bf SEFD} \\
                & {\bf (m)}      &                    & {\bf (Jy)}             &
                       & {\bf (Jy)}             \\ [2mm] \hline
    Culgoora      & 22 & RCP & 1550 &        &      \\
    DSS~43        & 70 &     &      & Linear &  400 \\
    Hartebesthoek & 26 & RCP & 1400 & RCP    & 1400 \\
    Hobart        & 26 & RCP & 1800 & RCP    & 4000 \\
    Mopra         & 22 & RCP & 1100 & Linear & 2200 \\
    Parkes        & 64 & RCP &  100 & RCP    &  140 \\ \hline
  \end{tabular} 
  \label{tab:vlbi_array}
\end{table}

\begin{table}
  {\scriptsize
  \caption[Coincident 6.7 and 12.2~GHz \methanol\/ masers in NGC~6334F]
          {Table of coincident 6.7 and 12.2~GHz \methanol\/ masers for
           NGC~6334F.}
  \begin{tabular}{rrrrrrrrrrr} \hline \hline
  \multicolumn{4}{c}{{\bf 6.7~GHz \methanol\/ masers}} & 
    \multicolumn{4}{c}{{\bf 12.2~GHz \methanol\/ masers}} &
    \multicolumn{2}{c}{{\bf Difference}} \\
  {\bf Velocity} & {\bf RA}     & {\bf Dec}    & {\bf Flux} &
    {\bf Velocity} & {\bf RA}     & {\bf Dec}    & {\bf Flux} &
    {\bf RA}       & {\bf Dec}      \\
  {\bf LSR}      & {\bf offset}   & {\bf offset}   & {\bf Density} &
    {\bf LSR}      & {\bf offset} & {\bf offset} & {\bf Density} &
    {\bf offset}   & {\bf offset}   \\
  {\bf (\kms\/)} & {\bf (arcsec)} & {\bf (arcsec)} & {\bf (Jy)}    &
    {\bf (\kms\/)} & {\bf (arcsec)} & {\bf (arcsec)} & {\bf (Jy)}    &
    {\bf (arcsec)} & {\bf (arcsec)} \\ [2mm] \hline
    -11.1 & 0.0000 &  0.0000 & 129 & -11.3 & 0.0000 &  0.0000 & 246 &  0.0000 &  0.0000 \\
    -10.6 & 2.2999 & -2.7332 &   9 & -10.8 & 2.2943 & -2.7346 &  14 &  0.0056 &  0.0014 \\
    -10.4 & 2.3168 & -2.3573 &  76 & -10.5 & 2.3192 & -2.3588 &  38 & -0.0024 &  0.0015 \\
     -9.9 & 2.4548 & -2.3049 &   4 &  -9.8 & 2.4536 & -2.3055 &   3 &  0.0012 &  0.0006 \\
     -9.0 & 2.5009 & -2.2602 &   4 &  -8.9 & 2.4951 & -2.2620 &   1 &  0.0058 &  0.0018 \\
  \end{tabular}
  }
  \label{tab:comp}
\end{table}

\newpage 

\end{document}